\documentstyle[editedvolume,numreferences,psfig]{crckapb} 
\newcommand{\avg}[1]{\langle{#1}\rangle}
\newcommand{\req}[1]{(\ref{#1})}
\newcommand{\beq}{\begin{equation}}
\newcommand{\eeq}{\end{equation}}
\newcommand{\beqas}{\begin{eqnarray*}}
\newcommand{\eeqas}{\end{eqnarray*}}
\newcommand{\beqar}{\begin{eqnarray}}
\newcommand{\eeqar}{\end{eqnarray}}

\begin{opening}
\title{STOCHASTIC DYNAMICS IN GAME THEORY}
\author{Matteo Marsili}
\author{Yi-Cheng Zhang}
\institute{Institut de Physique Th\'eorique, 
Universit\'e de Fribourg, CH-1700}
\end{opening}

\runningtitle{STOCHASTIC GAME DYNAMICS}
\begin{document}

\begin{abstract}
We introduce a simple stochastic dynamics for game theory. 
It assumes ``local'' rationality in the sense that any player 
climbs the gradient of his utility function in the presence of
a stochastic force which represents deviation from rationality 
in the form of a ``heat bath''. We focus on particular games 
of a large number of players with a global interaction which is 
typical of economic systems.
The stable states of this dynamics coincide with the Nash
equilibria of game theory.
We study the gaussian fluctuations around these equilibria
and establish that fluctuations around 
competitive equilibria increase with the number of 
players. In other words, competitive equilibria
are characterized by very broad and uneven distributions
among players. We also develop a small noise expansion
which allows to compute a ``free energy'' functional.
In particular we discuss the problem of equilibrium
selection when more than one equilibrium state is
present. 
\end{abstract}

\section{Introduction}

The economic world is a complex many-body dynamical system 
whose fluctuation phenomena has recently attracted much 
attention in the physicists community 
\cite{pw,scaling,multiscaling,soc,model}.
The identification, in economics and finance, of phenomena 
(such as scaling and multiscaling) which also occur
in physical systems (such as critical phase transitions 
and turbulence) suggests that some of the knowledge and 
techniques developed in physics to understand fluctuation
phenomena, might also be useful to understand
fluctuations in economics and finance.

Fluctuation phenomena in physics depend on the nature 
of the equilibrium state. The starting point to understand 
fluctuations in economics is then a theory for economic 
equilibria. Game Theory\cite{games} is the natural 
candidate: it describes 
the interaction among players' strategies and, assuming 
rationality, it identifies the possible game 
equilibria, named after Nash\cite{nash}.

It is important to stress at this point that we shall
deal mainly with economics and not with finance. 
Finance is, loosely speaking, a dynamical system out of 
equilibrium, where players gamble (speculate) on 
future market's fluctuations (see ref. \cite{model} for 
a model).
In an economic system, instead, the assumption
of rational behavior is more realistic. 
However perfect rationality is utopistic in
real life. Deviations from rationality, which can
arise from human errors, from limited or incomplete 
information or from random events, are 
practically unavoidable. Put differently
one can say that, in real life, reducing 
human errors or the impact of random events
costs time and money. Infinite precision is 
impossible if not at infinite costs.
The analysis of the effects of ``irrationality''
becomes then an important issue to understand how
game theoretical predictions can be modified in
realistic situations.

We shall address this issue for a ``thermodynamic''
economic system: a system with many degrees of
freedom (players). The random events discussed
above have then the same qualitative nature of
thermal fluctuations in statistical mechanics.
One can indeed think that, in a thermodynamic 
system, each degree of freedom pursues the 
minimization of the (global) energy in the 
presence of random shocks due to thermal 
fluctuations. In the same manner we shall 
assume that in a macro-economic system, each
agent pursues the maximization of his utility, 
under the effect of random shocks.
In this perspective, Nash equilibria 
become analogs of ground states in statistical 
mechanics and deviations from rationality can be 
introduced in exactly the same way as temperature 
is introduced in statistical mechanics.
In particular, there are several dynamical ways,
depending on the nature of the problem, 
to model temperature in statistical mechanics. 
This leads us to a natural definition
of stochastic dynamics in game theory.

In this work, which is an extension of a previous paper
\cite{fluct}, we shall follow these lines, using
the Langevin approach. We shall address
the issue of fluctuations around Nash equilibria
and the effects of deviations from rationality
in some specific games with a large number $n$ of 
players. 
We shall focus on games where each of the $n$ 
players can control a continuous variable or 
``strategy'' $x_i$. He is endowed of a utility 
function $u_i$ which depends on his strategy 
$x_i$ as well as on that of other players 
$\{x_j,~~j\neq i\}$.
In the model, each player continuously adjusts 
his variable $x_i$ in order to maximize his utility.
He also faces stochastic shocks which affect 
his actions and, as a result, 
the variable $x_i(t)$ becomes a continuous time 
stochastic process. The stochastic force acting on
a player is similar to that arising from an 
``heat bath'' at finite temperature in statistical 
mechanics.
The dynamics allows for a comparison with equilibrium
dynamics in statistical mechanics, which we find a
useful paradigm to discuss the results. In this comparison
negative utility plays the role of an energy and the effects
of fluctuations can be compared to entropic effects in
statistical mechanics. The main difference between the
two dynamics is that, while in statistical mechanics each
degree of freedom evolve to minimize a global (free) energy,
in game theory each degree of freedom maximizes his own
(part of the total) utility. 

\begin{figure}
\centerline{\psfig{file=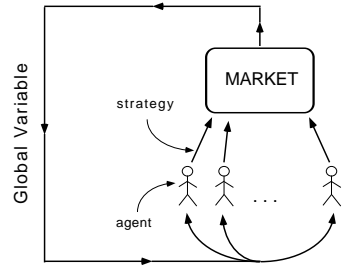,width=5cm,angle=0}}
%\vspace{5cm}
\caption{Global interaction among players 
in a stock market.}
\label{fig1}
\end{figure}

We shall focus on a particular class of games with
a {\em global interaction}. By this we mean that the
utility $u_i$ of each player depends on others' players 
strategies $x_j$ only through an aggregate or global
quantity $\bar x$, whose value is determined by all of 
them. This peculiar interaction, shown schematically
in figure \ref{fig1}, reflects the nature
of economic laws such as the law of demand and supply.

Nash equilibria are stationary points of the dynamics. 
When we include fluctuations we find that two main 
equilibria exist: {\em i)} non-competitive equilibria,
where each player's equilibrium strategy is determine by 
its interaction with the rules of the game and {\em ii)} 
competitive equilibria, which result from the 
the aggressive competition of each player with
all the others. For example, we shall discuss a game
where the introduction of taxes determines both a
non-competitive and a competitive equilibria. In the
former it will be the balance between profit and 
loss due to taxes, which is important for each 
player. In the other, competitive equilibrium,
taxes are negligible and the balance leading to
equilibrium is only due to the competition among
all players.

The main results that we shall find are:
\begin{enumerate}
\item Competitive Nash equilibria are usually affected by very large
fluctuations which increase with the number of players. 
Competition leads to broad distributions
and large inequalities in an economic system. This is 
reminiscent of Pareto Law of distribution of incomes\cite{pareto}.
We shall find that inequalities increase
with the number of players.
\item Competitive equilibria are also characterized by large relaxation
times which are proportional to $n$, and by a negative correlation 
between players' strategies. This means that two players tend 
to have opposite fluctuations around the Nash equilibrium.
\item At odd with statistical mechanics, where fluctuations
always increase the system's energy, we shall
find that in a game theoretical system, under particular 
conditions, fluctuations increase the utility (i.e. 
decrease the ``energy'').
\item We can, in principle, compute the stationary state
distribution in strategy space. This allows to solve, for
example, the problem of equilibrium selection in games where
more than one Nash equilibrium exists.
\item Fluctuations, in general, displace Nash equilibria and 
in some cases, for strong enough randomness, a Nash equilibrium 
can also disappear. 
\item Our approach also suggests that time-scales for the transition
from one Nash equilibrium to another one are proportional to 
$\exp[-\Delta F/D]$, where $\Delta F$ is a ``free energy barrier'' 
and $D$ is the noise strength.
\end{enumerate}

The paper is organized as follows. The next section
reviews game theory and discusses some simple game.
We also discuss briefly evolutionary game theory and
its differences with our approach.
Section \ref{model} introduces the class of models
we shall analyze. The following section discusses
gaussian fluctuations around Nash equilibria. Then
we develop a general approach to the stationary state
probability distribution in strategy space.
The main results are illustrated with
simple examples. 
In the final section we summarize the results, we 
draw some conclusions and comment on possible
further extensions.

\section{Game theory and evolution}

An economic system consists of a large number of 
interacting agents. In the game theoretical setting, 
each agent has a spectrum of strategies parametrized by 
an index $x$.
%, which can be a discrete as well as a continuous set.
Each player $i$ is also endowed of a utility or payoff
function $u_i$ which depends on the strategy $x_i$
he plays as well as on that played by all other players. 
With the notation $x_{-i}=\{x_j,~j\neq i\}$, we
can conveniently write $u_i=u_i(x_i,x_{-i})$. 
$x_i$ are also called {\em pure} strategy as
opposed to {\em mixed} strategies $\mu_i(x)$,
in which strategy $x$ is played with probability
$\mu_i(x)$ by player $i$.
Under mixed strategies, the strategies
used by the players are {\em independent} random 
variables. Independence is justified in one {\em stage}
games, which are played just once and each player has 
to decide his strategy simultaneously, without
information about what others will do.

Game theory, in its simplest setting, assumes that 
the payoff functions are common knowledge and that each
player behaves rationally. Rationality is also common
knowledge, which means that each player knows all
other players are rational (these are so-called
{\em complete information} games). 
Game theory aims at predicting the possible stable
states of the system, which are called {\em Nash
Equilibria} \cite{nash}. 
The strategies $x_i^*$ are a Nash Equilibrium (NE) if
each player's utility, for fixed opponents' strategies 
$x_{-i}^*$, is a maximum for $x_i=x_i^*$, i.e.  
$u_i(x_i^*,x_{-i}^*)\ge u_i(x_i,x_{-i}^*)$, $\forall x_i$.
The NE strategies $x_i^*$ are such that
no player has incentives to change his strategy,
since any change would cause a utility loss.
Nash showed\cite{nash} that any game has at least one 
Nash equilibrium in the space of mixed strategies.

\subsection{The Cournot Game and the Tragedy of the Commons}

The concept of NE is best illustrated by a simple 
example, originally introduced by Cournot in 1838 
\cite{courn}:
2 firms produce quantities $x_1$ and $x_2$ respectively, 
of a homogeneous product. The market-clearing price of 
the product depends, through the law of demand-and-offer,
on the total quantity $X=x_1+x_2$ produced: $P(X)=a-bX$.
The larger $X$ the smaller $P$ is. The model assumes
that the cost of producing a quantity $x_i$ is $cx_i$
and $c<a$. The firms choose their strategies (i.e. 
$x_i$) with the goal to maximize their 
profit. We can then define the payoff function
as $u_i=x_i[P(x_1+x_2)-c]=x_i[a-c-b(x_1+x_2)]$. 
The problem is to find $x_i$ assuming that both 
firms behave rationally. One way to do this is by means of
the concept of {\em best response}. The best response of a 
player $i$ to the opponents strategies $x_{-i}$, is the
strategy $x_i=x_i^*(x_{-i})$ which 
maximizes $u_i(x_i,x_{-i})$.
In the Cournot game, 
the best response $x_1^*(x_2)$ of firm $1$ to any given
strategy $x_2$ of firm $2$ is obtained by maximizing 
$u_1(x_1,x_2)$ with respect to $x_1$ with fixed $x_2$,
i.e. $x_1^*(x_2)=(a-c-bx_2)/(2b)$.
Firm $2$, knowing that $1$ behaves rationally (i.e. 
that it will play $x_1^*(x_2)$ whatever $x_2$ is) 
will choose $x_2^*$ which maximizes 
$u_2(x_1^*(x_2),x_2)$. It is important to stress
that operationally this means 
\[\left.\frac{\partial}{\partial x_2}u_2(x_1,x_2)
\right|_{x_1=x_1^*(x_2)}=0,\]
i.e. the maximum of $u_2$ must be found at {\em fixed}
$x_1$. This leads to $x_1^*=x_2^*=(a-c)/3b$. 
This simple example shows that rationality, and the
assumption of other's rationality, plays a crucial
role in the concept of Nash equilibrium\cite{nash}.
It is easy to generalize this game to $n$ firms.
Let us set, for convenience, $a-c=1$ and $b=1/n$. Then
$u_i(x_i,x_{-i})=x_i(1-\bar x)$, where $\bar x=(x_1+\ldots +x_n)/n$
is the average of $x_i$. The calculations generalizes 
straightforwardly and we find \cite{fluct} a NE at 
$x_i=x_0=n/(n+1)$ and a per player payoff $u_i=n/(n+1)^2$.
This celebrated example, is also known as the {\em 
Tragedy of the commons}\cite{hardin}. 
In this formulation of the problem,
the utility $u_i=x_i V(X)$ depends on a common resource
$V(X)=c-P(X)$ which, at the NE, is almost totally depleted 
by the aggressive behavior of players. As a result each player
receives a very small payoff. This is an example of a 
{\em competitive} NE where the strategies of players are
not limited because of a direct loss in utility, but
because the global resource is almost exhausted
by the aggressive, competitive behavior of players.

\subsection{Repeated Games and Evolutionary Game Theory}

This setting generalizes to repeated games, where 
different stage games are played a finite or an
infinite number of times. 
In repeated games a strategy must describe 
the action the player has to do at each stage.
Also the single stage utility is generally
replaced by an utility which accounts for 
many stages, usually with a {\em discount} factor
(i.e. an exponential weight function for future
utility). All these things make the analysis
much more complex than in single stage games.

A game posses, in general, several NE's, and this 
raises the question of which of them is more 
relevant. In order to answer this question, 
several definitions of stability have been proposed
\cite{evol,trembl,wujiang}.
The most successful approach to stability
has been that of evolutionary game theory\cite{maynard,evol}. 
This considers a game with mixed strategies 
as a game played by a population
of players playing pure strategies with random opponents.
%The fraction of players $i$ playing pure strategy 
%$x$ being the probability $\mu_i(x)$ of that pure strategy 
%in the mixed strategy. 
%One can analyze the stability 
%of a NE with respect to the introduction
%of a small share of ``mutant'' players, playing 
%with different strategies. If the newcomers' mutant
%population has a lower utility than that of players
%playing the NE strategies, the NE is said to be 
%{\em evolutionarily} stable.

This idea has attracted much interests in the community
of theoretical population biologists, which have translated
this idea into a mathematical model: the so called
{\em replicator dynamics}\cite{repl}. 
Though several versions of this dynamics have
been proposed \cite{repl,evol}, in its simplest 
form, it assumes that the individuals playing 
a given strategy reproduce at a rate proportional to 
their utility. 
Stochastic fluctuations, in the population
dynamical setting of replicator dynamics, have also
been considered in refs. \cite{FYFH}. 

There are some points that are worth to point out
in evolutionary game theory. The first is that its
application has been mostly limited to two players
games. Indeed its generalization to contexts with
$n$ players is technically very complex\cite{evol}.
The second is that rationality is {\em not} assumed. 
Players are on the contrary rather dull: they just 
keep playing the game the way their ancestors did. 
Rational NE results from the 
selective evolution of replicator dynamics. 
Finally we note that replicator dynamics 
assumes that the strategies $x_i$ are {\em independently}
chosen by each player\footnote{The joint distribution 
of the strategies factorizes into the single players
distribution functions $\mu_i(x,t)$.}. 
It also assumes that $x_i(t)$ is independent of 
$x_i(t')$ for $t'\neq t$.

In contrast with these observations, our goal is
to study a simple realistic dynamics for an $n\gg 1$
players game with ``almost'' rational players.
We shall do this weakening the assumption
of perfect rationality with the introduction of
a ``thermal'' noise. Therefore $x_i(t)$ will become a 
continuous time stochastic process.
We shall therefore pursue quite different purposes
and use totally different techniques than those
of evolutionary game dynamics.

\section{The Langevin approach}

Focusing on a class of models with a continuum spectrum of
strategies $x_i$, we recently proposed \cite{fluct}
a Langevin dynamics of the form
\beq
\partial_t x_i=\Gamma_i\frac{\partial u_i}{\partial 
x_i}+\eta_i\,.
\label{lang}
\eeq
Where the stochastic term $\eta_i(t)$ models deviation
from perfect rationality. We take $\eta_i(t)$ gaussian
with $\langle\eta_i(t)\rangle=0$ and 
\beq
\langle\eta_i(t)\eta_j(t')\rangle=2\Delta_{i,j}\delta(t-t').
\eeq

Eq. \req{lang} is a model dynamics which contains 
both the deterministic efforts each player exerts  to increase his
payoff and the effects of random events. The deterministic
part assumes ``local rationality'' of the agents: each agent
knows which is the direction in which his utility increases.
In other words, it assumes that each agent knows the utility
function $u_i$ as a function of $x_i$ in a small neighborhood of
$x_i$. Note that this weakens the assumption of
perfect rationality, according to which each player knows 
the function $u_i$ for any $x_i$ and any $x_j$, $j=1,\ldots,n$.

The stochastic term $\eta_i$, represents all hindrances which
prevent a rational behavior. These may be internal, i.e. 
affect only one player (e.g. illness), or external if they 
affect equally all players (e.g. earthquake). 
This suggests that $\eta_i$
is composed of two components, $\eta_i=\tilde\eta_i+\tilde\eta_0$,
where the $\tilde\eta_i$'s are independent gaussian 
forces. This motivates our choice 
\beq
\Delta_{i,j}=D_i\delta_{i,j}+D_0
\label{correl}
\eeq
for the correlation of $\eta_i$. 
Here $D_i$ is the strength of the stochastic force 
$\tilde\eta_i$ acting on player $i$, whereas $D_0$ is that for 
events $\tilde\eta_0$ which affect all players in the same way.

Clearly, since NE are defined as the set of $x_i$ for 
which the equations
\beq
\frac{\partial}{\partial x_i}u_i(x_i,x_{-i})=0,
\eeq
are simultaneously satisfied, for $\Delta_{i,j}=0$,
NE are stationary points of the dynamics. 
Equation \req{lang} is also very appealing since, 
comparing eq. \req{lang} with model A dynamics\cite{ref}, it
allows for an analogy with statistical mechanics.
The main difference with statistical
mechanics is that in game theory each degree
of freedom (player's strategy) maximizes a different
function (player's utility) whereas in statistical
mechanics each degree of freedom minimizes the same
function (energy or Hamiltonian). Also, at
odd with statistical mechanics, 
the interactions among strategies need not be symmetric: 
players' goals may be in conflict with one another. 

In this analogy, NE are analogs of zero temperature
(meta) stable states. 
The Langevin approach includes ``thermal'' fluctuations in
game theory, and this allows to analyze stability of
NE through the study of fluctuations. It also gives
a ``free energy'' measure, which enables to distinguish
the real ``ground state'' from ``meta-stable'' states.
Indeed in an ideal slow cooling down where $\Delta_{i,j}\to 0$, 
analogous to simulated annealing, only the state
with the smallest ``free energy'' is selected
{\em independently of the initial conditions}.
This contrasts with the evolutionary approach,
where the final state is uniquely determined 
by the initial conditions. 
The Fokker-Planck equation associated to eq. \req{lang}
provides a description of the game at the level of 
the distribution of $x_i$. At odd
with replicator dynamics (which also involves the distribution
functions of the $x_i$), this does not assumes that the $x_i$ are 
independent. As we shall see, correlations 
indeed arise.

\section{The model}
\label{model}

Many complex systems in economics have a very peculiar
form of interaction (see figure \ref{fig1}). 
In a stock market, for example,
each agent guesses whether to buy or to sell a stock,
looking at the stock's price fluctuations. These 
fluctuations are in their turn produced by the 
cumulative effect of the actions of all the agents
in the market, through some form of demand-offer law
\cite{model}. Each player interacts with a signal, which
in its turn is determined by the collective effect of
all players. A further example is the above mentioned
Cournot model, where $n$ firms produce the same good, 
and the market clearing price is determined by the ratio 
between the aggregate production and the demand. 

Focusing on this kind of interactions, we consider
in the following situations where the payoff 
function for player $i$ is
\beq
u_i(x_i,x_{-i})=-B(x_i,\bar x),~~~~~\bar x =
\frac{x_1+\ldots+x_n}{n}.
\label{payoff}
\eeq
In other words, the payoff for player $i$ depends on $x_j$
for $j\neq i$ only through the {\em aggregate} quantity
$n \bar x$. The $n$-players Cournot game, is of 
this form with $-B(x,y)=xV(y)$, and has been discussed at 
length in ref. \cite{fluct}. 

Because of the symmetry of the interaction, the NE are
symmetric: $x_i^*=x^*$ for all $i$, and $x^*$ satisfies
the equation
\beq
\left.\frac{\partial u_i}{\partial x_i}\right|_{x_j=x^*}=
-B_{1,0}(x^*,x^*)-\frac{1}{n}B_{0,1}(x^*, x^*)=0
\label{duidxi}
\eeq
where we defined, for convenience,
\[B_{j,k}(x,y)=
\frac{\partial^j}{\partial x^j}
\frac{\partial^k}{\partial y^k} B(x,y),
~~~B_{0,0}(x,y)=B(x,y).\]
A NE must not only be an extremum of $u_i$, it must also be
a maximum with respect to $x_i$ at fixed opponents'
strategies $x_{-i}=x^*$. This requires
\beq
\left.\frac{\partial^2 u_i}{\partial x_i^2}\right|_{x_j=x^*}=
-B_{2,0}(x^*,x^*)-\frac{n+1}{n^2}B_{1,1}(x^*, x^*)-
\frac{1}{n^2}B_{0,2}(x^*, x^*)<0.
\label{stability}
\eeq
It is worth to emphasize that eq. (\ref{duidxi},\ref{stability}) 
are necessary but not sufficient for $x^*$ to be a NE. 
Indeed a NE must be globally stable, which means that 
$u_i(x_i,x_{-i}=x^*)$
must have a global minimum at $x_i=x^*,~~\forall i$.

\section{Fluctuations around a Nash equilibrium}

Let $x_i=x^*$ be a NE for our model. Without loss of
generality we can set $x^*=0$ by a linear transformation
$x_i\to x_i-x^*$. We shall also set $B(0,0)=0$.
We can then investigate the small
gaussian fluctuations around the NE resulting
from the Langevin dynamics \req{lang}. Expanding
the deterministic part to leading order, we 
arrive at the equation
\beq
\dot x_i=-\Gamma_i\sum_{j=1}^n\left(g \delta_{i,j} +
\frac{h}{n} \right)x_j+\eta_i
\label{langq}
\eeq
where
\beqas
g&=&B_{2,0}(0,0)+\frac{1}{n}B_{1,1}(0,0),\\
h&=&B_{1,1}(0,0)+\frac{1}{n}B_{0,2}(0,0).
\eeqas
Stability requires that all the eigenvalues 
of the matrix $G_{i,j}=\Gamma_i(g \delta_{i,j} + 
{h}/{n})$ must be positive. These are given by 
the equation
\beq
\frac{1}{h}=\frac{1}{n}\sum_{k=1}^n\frac{\Gamma_k}{\lambda-
g\Gamma_k}.
\label{eigenv}
\eeq
A graphic inspection of the solutions of these
equations, shows that if
\beq
g>0,~~~~~~~h+g>0,
\label{condgh}
\eeq
then all eigenvalues are positive.
Note that, in terms of $g$ and $h$ the local
stability condition \req{stability} reads 
$g+h/n>0$. For $n\ge 1$ this is clearly satisfied
if the conditions \req{condgh} are met.

Equation \req{langq} is a multivariate 
Ornstein-Uhlenbeck process \cite{gardiner}. 
Fluctuations around the NE are described by the
matrix $\sigma_{i,j}=\avg{x_i x_j}$ of correlations.
This is the solution of the set of linear equations
$G\sigma+\sigma G^T=2\Delta$, where 
$G_{i,j}=\Gamma_i(g \delta_{i,j} + {h}/{n})$
and $\Delta$ is the matrix of the noise correlation
given in eq. \req{correl}\cite{gardiner}.
Introducing the vector $v_i=\sum_j\sigma_{i,j}$,
this matrix equation can be reduced to
\beqar
g\sigma_{i,i}+\frac{h}{n}v_i&=&\frac{D_i+D_0}{\Gamma_i}
\label{sii}\\
\left(g+h\overline{\frac{\Gamma}{\Gamma_i+\Gamma}}\right)v_i
+h\overline{\frac{\Gamma_i v}{\Gamma_i+\Gamma}}&=&
\frac{D_i}{\Gamma_i}+\overline{\frac{2nD_0}{\Gamma_i+\Gamma}}.
\label{eqv}
\eeqar
Here we have introduced the notation 
$\overline{f}=\frac{1}{n}\sum_kf_k$
for averages over the ensemble of players.  
Note that $v_i=n\avg{x_i\bar x}$ is the correlation between 
the variable $x_i$ and the global variable $\bar x$.

Qualitatively there are two different cases according
to whether the $\Gamma_i$ are broadly distributed or not.
We shall first focus on the second case, when the average 
value of $\Gamma_i$ is much larger than the fluctuations 
around it: $\overline{(\Gamma -\overline{\Gamma})^2}\ll 
\overline{\Gamma}^2$. With a redefinition of the scale of $B$,
we set, for simplicity, $\overline{\Gamma}=1$. 
In the limit $n\to\infty$
and $\epsilon=\sqrt{\overline{(\Gamma -\overline{\Gamma})^2}}\ll 1$,
the values of $\Gamma_i$ are densely distributed in a 
small interval of size $\approx \epsilon$. In each 
interval $[\Gamma,\Gamma+d\Gamma)$, $d\Gamma\ll\epsilon$ 
we can define an average value of $D_i$, 
$\sigma_{i,i}$ and $v_i$, which we denote by 
$D(\Gamma)$, $\sigma(\Gamma)$ and $v(\Gamma)$. 
This allows for a systematic expansion in powers of
$\epsilon$. 

For $\epsilon=0$, one easily finds
\beqar
v(1)=\overline{v}&=&\frac{\overline{D}+nD_0}{g+h},\nonumber\\
\sigma(1)=\overline{\sigma}&=&
\left(1-\frac{h}{n(g+h)}\right)\frac{\overline{D}}{g}+
\frac{D_0}{g+h}
\label{s0}
\eeqar
Note that $v(1)>0$, which means that each 
variable $x_i$ tends to fluctuate in phase with 
$\bar x$.
We can also compute an ensemble average of the
correlation, which for $\epsilon=0$ reads
\beq
C=\frac{1}{n(n-1)}\sum_{i\neq j}\avg{x_i x_j}=
\frac{\overline{v-\sigma}}{n-1}=
\frac{D_0}{g+h}-\frac{h\overline{D}}{ng(g+h)}.
\label{ccorrel}
\eeq
The common stochastic force, as it could be expected, gives 
a positive contribution to the correlation, and for $D_0$
large enough, the correlation always turns positive.

With respect to the dynamics in the stationary state,
correlation functions decay exponentially 
\[\avg{x_i(t+t_0)x_j(t_0)}-\sigma_{i,j}\propto e^{-t/\tau}\,.\]
The correlation time $\tau$ of the leading exponential 
behavior is given by the minimum
eigenvalue $\tau =\max_k(\lambda_k^{-1})$ in eq. \req{eigenv}.

It is finally possible to compute the average utility
\beq
\overline{\avg{u}}=-
\frac{1}{2}\overline{\sigma}B_{2,0}-
\frac{\overline{v}}{n}B_{1,1}-
\frac{\overline{v}}{2n}B_{0,2}
\label{avgu}
\eeq
where all derivatives of $B$ are evaluated at $(0,0)$.
Note that in view of our choice $B(0,0)=0$ this expression
yields the deviation of the average utility from a completely
rational behavior. As we shall see it is possible that fluctuations
increase the average utility.
The last term in eq. \req{avgu} is the average
of the utility {\em at} the Nash equilibrium 
{\em in presence} of fluctuations: 
\[\avg{u_i(0,x_{-i})}=-\frac{\overline{v}}{2n}B_{0,2}.\]
This would be the utility of a player which maintains the
NE strategy $x_i=0$ even in presence of fluctuations.
It is interesting to note that it is possible that
$\overline{\avg{u}}>\avg{u_i(0,x_{-i})}$. Loosely
speaking this means that in a game with random deviations
from rationality players who are affected by the
randomness may receive a higher payoff (on average)
than those which behave rationally ($x_i=0$).
The condition for this to happen is 
\beq
\frac{1}{2}\overline{\sigma}B_{2,0}+
\frac{\overline{v}}{n}B_{1,1}<0\,.
\label{condirr}
\eeq

\begin{figure}
\centerline{\psfig{file=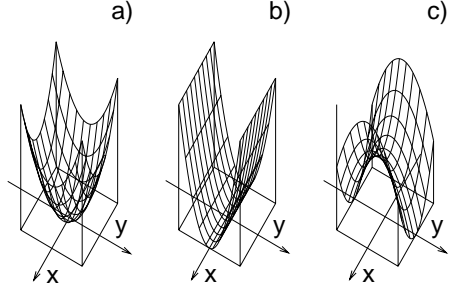,width=7cm,angle=270}}
%\vspace{5cm}
\caption{Quadratic approximation of the utility function close
to a Nash equilibrium. a) $B_{2,0},B_{1,1}$ and $B_{2,0}$ are 
finite and positive; b) $B_{2,0}\simeq 0$, $B_{1,1}>0$
 and $B_{2,0}>0$; c) $B_{2,0}+B_{1,1}\simeq 0$.}
\label{fig2}
\end{figure}

Let us now discuss these findings. As expected the 
fluctuations of $x_i$ grow with $D_i$ and $D_0$. 
There are three qualitatively different cases, as shown
in figure \ref{fig2}:
\begin{description}
\item[ a)] when $B_{2,0},B_{1,1}$ and $B_{2,0}$ are 
all finite and positive. The point $(g,h)$ lies well 
inside the domain defined by eq. \req{condgh}. 
As a result we have a normal behavior with small 
fluctuation. Fluctuations increase the average utility
and a rational behavior is generally more rewarding.
\item[ b)] $B_{2,0}\simeq 0$, $B_{1,1}$
 and $B_{2,0}$ are finite and positive. Then 
$g\sim 1/n$ is small and, from eq. \req{s0}, we see that 
fluctuations are proportional to $n$. 
This, in view of the explicit factor $g$ in front of 
$\sigma_{i,i}$ in eq. \req{sii}, is a 
general feature which holds also for broadly distributed 
$\Gamma_i$. The condition $B_{2,0}\simeq 0$ 
obtains for example for utilities of the form
$B(x,y)=-x b(y)$, which describes e.g. the tragedy of the 
commons problem\cite{hardin,fluct}. We shall discuss in more
detail this class of models in the next paragraph.
Generally $B_{2,0}=0$ is typical of competitive equilibria.
Indeed it means that each player  does not feel 
the effects of a change in his $x_i$ directly. Rather it feels 
it indirectly through the reaction of other players, or
better through the effects this change has on the global
variable $\bar x$. Large fluctuations are a result of
the fact that the dependence of $\bar x$ on $x_i$ is weak.
Competitive equilibria are also characterized by negatively
correlated variables $x_i$ for $D_0$ small enough: $C<0$.
Large fluctuations come together
with large relaxation times.
Indeed eigenvalues are proportional to $g$, 
so that for $g\ll 1$ all of them are small, yielding
large relaxation times $\tau\sim n$.
Finally the average utility is decreased. 

\item[ c)] $B_{2,0}+B_{1,1}\simeq 0$. Also in this case 
large fluctuations occur since  $g+h\simeq 0$. 
The divergence of the terms proportional to $D_0$ suggests
that the mode with stronger fluctuations is 
associated with $\bar x$. For $h+g$ small the 
smallest eigenvalue is small. This results in a large
correlation time $\tau=\overline{\Gamma^{-1}}/(g+h)+O(1)$.
At odd with the case {\em b)},  only one 
eigenvalue is small in this case, the others being 
$O(1)$. The (right) eigenvector associated with this 
eigenvalue is $w_i=1+O(g+h)$ nearly constant.
The slow mode characterized by large fluctuations is then
associated with $\bar x$.
Note that also $C\sim (g+h)^{-1}$ is large and 
positive, which means that $x_i$ fluctuate in phase
thus yielding a large fluctuation of their sum.
Finally fluctuations may {\em decrease} the average utility 
in this case. Furthermore if player
$i$ behaves rationally ($x_i=0$) he receives a smaller payoff
$\overline{\avg{u}}>\avg{u_i(0,x_{-i})}$.
\end{description}

These results can be extended to higher order in $\epsilon$.
The idea is to assume that $\Gamma_i$ are distributed 
around $1$ according to a gaussian density with standard 
deviation $\epsilon$. We shall limit our discussion to
the first term here. Taking the average of eq. \req{eqv},
multiplied by $\Gamma_i-1$ over this distribution and
taking the leading order in $\epsilon$, gives
\beqas
v'(1)&=&-\frac{D(1)}{g+h}+\frac{2D'(1)}{2g+h}-
\frac{gnD_0}{(g+h)(2g+h)}\\
\sigma'(1)&=&\frac{D'(1)-D(1)-D_0}{g}-\frac{h}{gn}v'(1)\,.
\eeqas
These equations give interesting informations. For 
example $\sigma'(1)<0$ implies that the fluctuations 
experienced by a player are smaller the faster his 
dynamics (i.e. the larger his rate constant $\Gamma$). 
This is what one naturally
expects and it occurs when $D'(1)<D(1)+D_0
[1-gh/(g+h)(2g+h)]+O(1/n)$. On the other hand, if 
$D(\Gamma)$ grows sufficiently fast with $\Gamma$, 
one has $\sigma'(1)>0$. This suggests that 
generally the fluctuations of a variable $x_i$ grow 
with $D_i$ and decrease with $\Gamma_i$.
The same kind of information can 
be obtained for the correlation $C$. 
%\[C'=\frac{hD}{ng(g+h)}-\frac{hD'}{ng(2g+h)}
%-\frac{nD_0}{(g+h)(2g+h)}\left[g-\frac{g^2+(g+h)^2}{ng}\right]\]
The case $D_0=0$ yields a compact expression:
\[\frac{C'}{C}=-1+\frac{g+h}{2g+h}\frac{D'}{D}.\]
This says that correlations are weaker for faster
variables, unless $D(\Gamma)$ grows fast enough
with $\Gamma$. We shall not discuss the case $D_0>0$,
which leads to less transparent formulas.

The case of broadly distributed $\Gamma_i$ needs a
more detailed knowledge of the parameters. However
we expect that the results obtained by the small
disorder expansion above qualitatively describe 
the system. Note that eq. \req{sii} suggests that
$\avg{x_i^2}\propto D_i/\Gamma_i$ diverges as 
$\Gamma_i\to 0$. In a large system of players with 
broadly distributed $\Gamma_i$, the smallest 
$\Gamma_i$ can be vanishingly small as $n\to\infty$.
For example, in a system where the
$\Gamma_i$ are distributed with a density 
$\rho(\Gamma)\sim \Gamma^{\beta-1}$ for $\Gamma\ll 1$,
one expects that the smallest $\Gamma_i$, in the population 
of players, is $\Gamma_{\min}\sim n^{-1/\beta}$. 
In this case some player can have fluctuations growing
with $n$. It is worth to stress, however that such a 
distribution of $\Gamma_i$ implies a power law distribution 
of characteristic times $1/\Gamma_i$, which might not
be realistic. 

\subsection{An example}

Let us illustrate these findings with simple
examples. An example with an utility of the form
$B(x,y)=xy$ has been discussed in detail
elsewhere \cite{fluct}. The main point raised
by this example is that of the emergence of 
large fluctuations. Here we shall describe a
second example:
\[B(x,y|\rho)=\frac{b}{2}(x-\rho y)^2.\]
The utility function $u_i(x_i,x_{-i})=-B(x_i,\bar x|\rho)$
above describes a classical game where each player has to
throw a number $x_i$ with the aim of guessing a 
fraction $\rho$ of the average $\bar x$ of all players' guesses.
This clearly has only one NE $x_i=0$, $\forall i$.
$B(x,y|\rho)$ can also be assumed as a local 
approximation of a more complex utility around one
of its Nash equilibria.

With the choice $b=(1-\rho/n)^{-1}$, the parameters
are $g=1$ and $h=-\rho$. The stability condition 
requires that $\rho <1$, which is intuitive because
if players where told to guess a number larger than the
mean everybody would tend to overshoot and $x_i\to
\infty$. On the contrary with $\rho<1$ every player 
has to be careful: he must play a number which is
smaller than the one played by the others.
In the extreme case $\rho<0$ he has to try to do
the opposite of what the majority does. Let us 
discuss only the results for $\epsilon=0$.
It is straightforward to find
\beqas
\sigma&=&\overline{D}\left(1+\frac{\rho}{n(1-\rho)}\right)
+\frac{D_0}{1-\rho}\\
C&=&\frac{\rho \overline{D}}{n(1-\rho)}+\frac{D_0}{1-\rho}
\eeqas
Note that, for $D_0=0$, $C$ has the same sign of 
$\rho$. For $\rho>0$, a player attempts to guess the
fluctuation of others and as a result it tends to 
make fluctuations in the same direction as the others.
On the other hand, for $\rho<0$ a negative correlation
arises.

The average utility is $\overline{\avg{u}}=-\frac{1}{2}
\overline{D}-\frac{n}{2}\frac{1-\rho}{n-\rho}D_0$ whereas
the condition \req{condirr}, after some algebra, reads:
\[[n-(n+1)\rho]\overline{D}+(1-2\rho)D_0< 0\,.\]
In order for this condition to hold at least one of
the two terms must be negative. For 
$\frac{1}{2}<\rho<\frac{n}{n+1}$, it becomes 
``favorable'' to follow the random force for 
\[D_0\geq \frac{n-(n+1)\rho}{n(2\rho-1)}\overline{D}\,.\]
In other words, if the global component of the 
stochastic force is strong enough, it is not 
convenient to play the NE strategy $x_i=0$.

This behavior can be qualitatively explained as follows:
Each player has to try to follow as closely as 
possible the global variable $\rho\bar x$. The latter evolves
under a stochastic force $\bar\eta$ of strength 
$\overline{D}/n+D_0\simeq D_0$. The random force $\eta_i$
acting on each player has a component of strength 
$D_0$ along $\rho\bar x$. If this component is large
enough, each player can guess correctly whether the 
mean $\bar x$ will move left or right and so it 
becomes favorable to follow the stochastic force.

For $\rho>\frac{n}{n+1}$ the condition holds $\forall 
\overline{D}>0$ with $D_0=0$. This region is also 
characterized by large correlated fluctuations 
(note that $g+h=1-\rho\sim 1/n$). Even in the absence 
of a global stochastic force, the dynamics leads
to a state where the $x_i$ are highly correlated.
In such a state, the strategy $x_i=0$ is less
rewarding than the average.

For $\rho<\frac{1}{2}$ there are no values of 
$\overline{D}$ and $D_0$ for which the condition
\req{condirr} is satisfied.

\section{Probability distribution in strategy space}

In this section we shall extend the analysis of
our model to study the full probability distribution
in the stationary state. A complete treatment is
not possible in general. We shall restrict
attention to the case
\[\Gamma_i=1,~~~~~D_i=D\,.\]
%and then discuss how the results can be extended to
%general $\Gamma_i$ and $D_i$. 
In view of our discussion of the previous section, 
$\Gamma_i=1$ means that all players have the same
characteristic time-scale. Qualitatively, we expect
that the results below apply also in case of narrowly 
distributed time-scales.

Our equation is
\beq
\dot x_i=-B_{1,0}(x_i,\bar x)-\frac{1}{n}B_{0,1}(x_i,\bar x)+\eta_i.
\label{eqxi}
\eeq

It is useful to introduce the variables
\beqar
y_k&=&\frac{1}{\sqrt{k(k+1)}}\sum_{i=1}^k(x_i-x_{k+1}),~~~~~~~~ k<n\\
y_n&=&\frac{x_1+\ldots+x_n}{\sqrt{n}}=\sqrt{n}\bar x.
\eeqar
The transformation $\vec x\to\vec y$ is orthonormal, 
which implies the useful identity
\beq
\sum_{i=1}^nx_i^2=\sum_{k=1}^n y_k^2.
\label{ident}
\eeq
The same transformation, applied to the noise term $\vec\eta\to
\vec\zeta$ leads to a stochastic force $\zeta_k$ in the equation for 
$\dot y_k$ which has a correlation
\beq
\avg{\zeta_j(t)\zeta_k(t')}=2\delta_{j,k}(D+nD_0\delta_{k,n})
\label{czeta}
\eeq
which is diagonal. The common 
stochastic force $D_0$ acts on $y_n$ only.
For convenience, instead of $y_n$, we shall use
the variable $\bar x=y_n/\sqrt{n}$. The noise $\bar \eta$ 
appearing in the equation for $\dot {\bar x}$ has a strength
$T=D/n+D_0$.

\subsection{Linear utility}

Let us first consider the model 
\[B(x,z)=x b(z),\]
which allows for a full solution for the stationary state
distribution of $x_i$. A situation described by this
kind of utility function \cite{hardin,fluct} is a system 
where $n$ firms produce a quantity $x_i$ of a homogeneous product.
Then $-b$ is the difference between the market clearing price of 
one unit of product and the production cost of one unit.
We assume that it depends only on the 
aggregate production $\sum_i x_i=n\bar x$ (the production cost 
per unit is a constant).
The payoff $u_i=-x_ib$ of firm $i$ is then proportional
to its production. In realistic situations $b(x)$ is an
increasing function. One expects e.g. that because of
competition, the price $-b$ of a product decreases 
with the total quantity produced, in view of the law
of demand and offer.

The NE $x_i=x^*$ is defined by $b(x^*)=-x^*b'(x^*)/n$. Note that
the payoff per player 
\[u_i=-x^*b(x^*)=\frac{{x^*}^2b'(x^*)}{n}\]
is positive and proportional to $1/n$. 

The orthonormal transformation $\vec x\to \vec y$, yields
\beqar
\dot y_k&=&-\frac{b'(\bar x)}{n} y_k+\zeta_k,~~~~~~~~k<n
\label{eqyk}\\
\dot {\bar x}&=&-b(\bar x)-\frac{b'(\bar x)}{n} \bar x+\bar \eta.
\label{eqyn}
\eeqar
The equation for $\bar x$ does not involve other variables
and can be directly solved yielding the distribution 
$p_n(\bar x)$. The equations for $y_k$ depend 
only on $\bar x$. Treating $\bar x$ as a parameter, one can 
find the conditional distribution of $y_k$: $p(y_k|\bar x)$.
The full distribution is then
\[p(y_1,\ldots,y_n)=p_n(\bar x)\prod_{k=1}^{n-1}p(y_k|\bar x).\]
Back transforming to the variables $x_i$, yields the solution.
Eq. (\ref{eqyn}) describes a ``particle'' in a potential with 
thermal fluctuations and can be solved using standard 
techniques\cite{gardiner}:
\beq
p_n(\bar x)={\cal N}\exp\left\{-\frac{{\bar x}b({\bar x})+
(n-1)\int_0^{\bar x}dz b(z)}{D+nD_0}\right\}
\label{pnbarx}
\eeq
with ${\cal N}$ a normalization factor.
The equation for $y_k$ similarly gives
\beq
p(y_k|\bar x)=\sqrt{\frac{b'(\bar x)}{\pi n}}
\exp\left\{-\frac{b'({\bar x})y_k^2}{nD}\right\}
\label{pyk}
\eeq
where here normalization requires $b'({\bar x})>0$.
Using the equation \req{ident}, one can easily find the full
distribution of $x_i$:
\beqas
p(x_1,\ldots,x_n)\propto\left[\frac{b'(\bar x)}{\pi nD}
\right]^{\frac{n-1}{2}}\exp\left\{-\frac{b'({\bar x})}{nD}
\sum_{i=1}^n(x_i-{\bar x})^2\right.&~&\\
-\left.\frac{{\bar x}b({\bar x})+
(n-1)\int_0^{\bar x}dz b(z)}{D+nD_0}\right\}&~&
\eeqas
where $\bar x=(x_1+\cdots+x_n)/n$.

Some implications of this result have already been discussed
in ref. \cite{fluct}. In particular it was observed 
that if $b'(x)\sim O(1)$ one finds fluctuations of order
$\avg{x_i^2}\propto nD$ and large relaxation times $\tau\sim n$. 
We note furthermore that $p(x_1,\ldots,x_n)$ vanishes as
$b(\bar x)\to 0^+$. For $b'(\bar x)<0$, which corresponds
in any case to an ``unphysical'' situation\footnote{For
example,  in the firms problem, $b'(\bar x)<0$
means that the price increases if the 
production is increased.}, 
we must set $p(x_1,\ldots,x_n)=0$.
Note that eqs. (\ref{eqyk}, \ref{eqyn}) imply that if the 
system is ``prepared'' at $t=0$ in a state with $b(\bar x)<0$,
in the early stages of the dynamics, the deviations
$x_i-\bar x$ increase exponentially. 
This is clearly a highly non-equilibrium 
situation. 

The average utility, to order $D+nD_0$ is given by
\beq
\avg{u_i}\simeq-x^*b(x^*)-\frac{[2b'(x^*)+x^*b''(x^*)](D+nD_0)}
{2[(n+1)b'(x^*)+x^*b''(x^*)]}\,.
\label{avgut}
\eeq
If $\frac{\partial^2}{\partial x^2}[ x b(x)]_{x=x^*}<0$, then
the effect of fluctuation will be that of increasing the average
utility (note that, in the notations of the
previous section, $(n+1)b'(x^*)+x^*b''(x^*)=n(g+h)>0$).
An example, which allows for simple expressions, is
$b(x)=-1+x-\frac{1}{2}ax^2$. Since $b'(x)=1-ax$ we need to restrict
the range of $x$ to $x<1/a$ in order for $b'(x)>0$ (otherwise
$\bar x$ would flow to $\infty$). The NE is at 
$x^*=\frac{n+1}{a(n+2)}\left(1-\sqrt{1-2a+\frac{2a}{(n+1)^2}}\right)
\simeq\frac{1}{a}(1-\sqrt{1-2a})$ and its existence
requires $a\le 1/2$. The payoff per player at the NE, 
to leading order in $n$, is 
$-x^*b(x^*)\simeq\frac{1}{na^2}(1-\sqrt{1-2a})^2\sqrt{1-2a}$. 
With gaussian fluctuations, we find:
\[\avg{u_i}\simeq -x^* b(x^*)\left[1-\frac{a^2(3\sqrt{1-2a}-1)
(D+nD_0)}{2(1-2a)(1-\sqrt{1-2a})^2}\right].\] 
For $a\ge 4/9$ fluctuations increase the average utility, 
an effect related to the asymmetry of the function
$b(x)$ around $x^*$. 

The average $\avg{\bar x}$ instead, has no corrections to
order $D$. As we shall see this does not hold in general.
It does neither hold when the function $b(x)$ changes rapidly
close to the NE, a situation which cannot be described in the
gaussian approximation. Consider for example, the game of the
tragedy of the commons, where the utility function 
is much steeper when negative payoff arise: $b(x)=x-1$ for
$x<1$ and $b(x)=q(x-1)$ for $x>1$ and $q\gg 1$. This represents
a situation where each player is very scared of receiving
negative payoffs. Clearly, as far as the NE is concerned,
no change occurs: The NE $x^*=n/(n+1)$ is always the same, 
dangerously close to the edge of negative utilities. In the presence
of fluctuations, however, the distribution of $\bar x$ is very 
asymmetric on the two sides of the NE. For $\bar x>x^*$ it drops off
much more quickly than for $\bar x<x^*$. As a result the
NE is shifted by an amount of order $\sqrt{D/n}$ towards 
safer values of $\bar x<x^*$. We see then that fluctuations can 
induce a more cautious behavior. 

The most general model 
which allows for a complete solution, with the above technique, 
is with $B(x,z)=B_0(z)+xb(z)+cx^2$. 
The term $B_0(z)$ changes only the 
distribution of $\bar x$ by a factor
proportional to $\exp[-B_0(\bar x)/n]$, whereas the term $cx^2$
also affects the distribution of $x_i$. 
A simple realization of 
this model, with $c=0$, is the one where players cooperate
to increase each other's utility: $B_0(z)=\beta zb(z)$ ($\beta>0$).
Of course $\beta<0$ means anti-cooperation, i.e. each 
player tries to maximize his utility as well as to minimize
that of others. 
With $b(z)=z-1$, one easily finds that
the NE is at $x^*=\frac{n+\beta}{n+2\beta+1}$ and 
the payoff per player is $u_i^*=\frac{(n+\beta)(1+\beta)}
{(n+2\beta+1)^2}$. A high degree of cooperation,
$\beta\propto n$ leads to a finite utility per player. 
On the contrary fluctuations always remain large 
$\avg{\delta x_i^2}\sim nD$. For $\beta\to\infty$, as
discussed in \cite{fluct}, fluctuations diverge
even though the average utility remains finite.
Clearly anti-cooperation $\beta<0$ decreases 
the utility. However for $\beta<-1$ fluctuations 
give a positive contribution to the utility. 
Fluctuations increase the average utility in 
over-competitive systems (those in which each player
main efforts are devoted to decrease other players'
payoffs).

\subsection{The general case}

The general case 
does not allow for a full solution. It is however
possible to compute the stationary state distribution to leading
order in $D$. 
We assume that 
\beq
y_k\sim x_i-\bar x\sim O(\sqrt{D}).
\label{assump}
\eeq
While this is surely satisfied close to a NE (when all $x_i$ 
are close to $x^*$), it might not hold in non-equilibrium 
situations or when rare events such as large fluctuations
occur.

The equation for $\dot y_k$, derived from eq. \req{eqxi},
contains terms $B_{j,k}(x_i,\bar x)-B_{j,k}(x_{k+1},\bar x)$
with $(j,k)=(1,0)$ or $(0,1)$. Expanding in powers of 
$x_i-\bar x$ and $x_{k+1}-\bar x$, we find, to leading
order
\[\dot y_k=
-\left[B_{2,0}(\bar x,\bar x)+\frac{1}{n}B_{1,1}(\bar x,\bar x)
\right]y_k+\zeta_k\]
where $\zeta_k$ is still gaussian uncorrelated noise,
in view of the orthonormality of the transformation $\vec x\to
\vec y$ (see eq. \ref{czeta}). 
This equation is valid to $O(D)$, since we neglected
terms proportional to $(x_i-\bar x)^2$ which are of order $D$
in view of eq. \req{assump}.
Using eq. \req{czeta}, one easily finds:
\beq
\avg{y_k^2|\bar x}=\frac{nD}
{nB_{2,0}(\bar x,\bar x)+B_{1,1}(\bar x,\bar x)}
\label{fluctyk}
\eeq
where we used the notation $\avg{y|x}$ for the 
average of the quantity $y$ conditional to the value
$x$ assumed by a second variable.
Note that the requirement $\avg{y_k^2|\bar x}\ge 0$, 
implies $nB_{2,0}(\bar x,\bar x)+B_{1,1}(\bar x,\bar x)\ge 0$.
This condition, which generalizes the condition $b'(\bar x)\ge 0$ in
the previous paragraph, restricts the range of possible values of 
$\bar x$.

Let us now move to the equation for $\bar x$. Expanding
the right hand side of the equation for $x_i$ to
second order in $x_i-\bar x$, we find
\beqas
\dot{\bar x}=&-&B_{1,0}(\bar x,\bar x)-
\frac{1}{n}B_{0,1}(\bar x,\bar x)\\
&-&\frac{1}{2}
\left[B_{3,0}(\bar x,\bar x)+
\frac{1}{n}B_{2,1}(\bar x,\bar x)\right]\frac{1}{n}\sum_{i=1}^n
(x_i-\bar x)^2+\ldots+\bar\eta
\eeqas
where $\avg{\bar\eta(t)\bar\eta(t')}=2(D/n+D_0)\delta(t-t')$.
In view of eq. \req{ident}, we have $\sum_{i=1}^n
(x_i-\bar x)^2=\sum_{k=1}^{n-1} y_k^2$. 
Taking the average over $y_k$ conditional to
the value of $\bar x$ in this equation, we 
substitute $\frac{1}{n}\sum_i(x_i-\bar x)^2$ with 
$\frac{n-1}{n}\avg{y_k^2|\bar x}$. This results in the equation
\[\dot{\bar x}=-B_{1,0}-
\frac{1}{n}B_{0,1}-\frac{(n-1)D}{2n}\frac{nB_{3,0}+B_{2,1}}
{nB_{2,0}+B_{1,1}}+\bar\eta\]
where we suppressed the dependence on $(\bar x,\bar x)$
of $B_{k,l}$.

The steady state distribution of $\bar x$ is then given
by $P(\bar x)\propto \exp[-F(\bar x)/T]$, where the 
``temperature'' is $T=D/n+D_0$, and the ``free energy''
is given by:
\beq
F(x)
%&=&\int^{x}dy\left\{B_{1,0}(y,y)+
%\frac{1}{n}B_{0,1}(y,y)+\frac{1}{2}\left[B_{3,0}(y,y)+
%\frac{1}{n}B_{2,1}(y,y)\right]\avg{y_k|\bar x =y}
%\right\}
\label{fx}
%\\
=\int^{x}\left[B_{1,0}+
\frac{1}{n}B_{0,1}+\frac{(n-1)D}{4n}\frac{nB_{3,0}+B_{2,1}}
{nB_{2,0}+B_{1,1}}\right]
%\nonumber
\eeq
It is useful, for the following discussion, to split
$F=U-DS$ into a $D$ independent part and into a 
$D$ dependent one:
\[U(x)=\int^{x}\left[B_{1,0}(y,y)+
\frac{1}{n}B_{0,1}(y,y)\right]dy\]
and
\[S(x)=-\frac{n-1}{2n}\int^{x}dy\frac{nB_{3,0}(y,y)+B_{2,1}(y,y)}
{nB_{2,0}(y,y)+B_{1,1}(y,y)}\]
The relation $F=U-DS$ is reminiscent of a free energy in 
equilibrium statistical mechanics, which is a useful 
paradigm to discuss our system.

\subsection{Discussion and applications}

It is worth to stress that the function $U(x)$ is not 
simply related to the utility. This contrast with equilibrium 
statistical mechanics where the probability of a state is directly 
related to his energy. 

The ``entropic'' term $S(x)$ results from the inclusion of the 
fluctuations of the variables $y_k$. Usually, in statistical 
mechanics, one finds that the larger the fluctuations in a 
state the larger the entropy is. We shall see in the following 
that this does not hold for our system: $S$ can be larger
for ``ordered'' states than for states with wild fluctuations.

Fluctuations displace the NE (defined as the minima
of $F(x)$) of a quantity of order $D$:
\[x^*(D)=x^*(0)-\frac{(n-1)D(nB_{3,0}+B_{2,1})}
{2n^2(nB_{2,0}+B_{1,1})[nB_{2,0}+(n+1)B_{1,1}+B_{2,0}]}.\]
Note that stability requires that the denominator be positive. 

The analysis of any particular case goes as follows.
First one needs to determine the range of $\bar x$
where our approach can be applied. This is given by the
condition $nB_{2,0}(\bar x,\bar x)+B_{1,1}(\bar x,\bar x)>0$
which ensures that $\avg{y_k^2|\bar x}>0$ is finite. 
Outside this range, the behavior cannot be described 
perturbatively in $D$ (i.e. the gaussian approximation for 
the variables $y_k$ is no longer valid).
Secondly find all solutions of 
eq. \req{duidxi}, $nB_{1,0}+B_{0,1}=0$, which
are the candidates for NE's. Each solution to this
equation must then be checked for stability. If 
eqs. \req{condgh} are verified, the equilibrium is
stable. Finally for each stable equilibrium one can 
evaluate its free energy $F(x^*)$ from eq. \req{fx}.
This gives the statistical weight of each state in the
stationary regime. The state with the smallest $F(x^*)$
is the one with a larger statistical weight and
it dominates in the limit $T=D/n+D_0\to 0$. 

In ref. \cite{fluct} we discussed the case 
\beq
B(x,y)=-x(1-y)\left[1-\frac{x(1-y)}{2c}\right]\,,~~~~~c>0
\label{b2}
\eeq
of a quadratic utility both in $x$ and $y$. This is an 
interesting case, both because such an utility function 
can be motivated \cite{fluct} and because the system possesses
two NE's. As a rough motivation, we can go back to the firms
problem with $b(y)=-1+y$, and argue that their utility $u_i$ 
may not exactly equal their net gain $x_i(1-\bar x)$ since
this is then subject to taxes. If taxes do not grow linearly
with the income (as is usually the case) we need to add a 
further term to the utility. The simplest choice leads to
the above form of $B(x,y)$.

Let us go through the above passages for this
example: Eq. \req{duidxi} has one only solution for $c>1/4$ 
which is at 
\beq
x_0=\frac{n}{n+1}\,,
\label{x0}
\eeq
which is stable $\forall c>0$.
For $c<1/4$ two other solutions appear at
\beq
x_{\pm}=\frac{1\pm\sqrt{1-4c}}{2}
\label{xpm}
\eeq
of which only $x_-$ is stable. These two NE's have a quite different 
nature. The NE at $x_0$ is a competitive
NE since $B_{2,0}\ll 1$ and it is characterized
by a small payoff per player $u_i\sim 1/n$ and by large 
fluctuations $\avg{\delta x_i^2}\sim n/D$. 
The NE at $x_-$ is {\em Pareto dominant}\footnote{The NE with 
largest utility is called {\em Pareto dominant} equilibrium in 
game theory.} since it has a finite and positive utility.
Also fluctuations are finite, as $n\to\infty$, at $x_-$.
At this NE the action of players is limited by the
increase of the non-linear term due to ``taxes''.

\begin{figure}
\centerline{\psfig{file=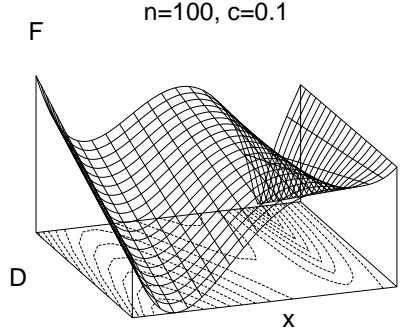,width=7cm,angle=270}}
%\vspace{5cm}
\caption{$F(x)$, as a function of $D$ for the 
quadratic utility function: $n=100$ and $c=0.1$.}
\label{fig3}
\end{figure}

The function $F(x)$ is readily computed.
Setting, for convenience $F(x_+)=0$, we find
\[F(x_-)=-\frac{(1-4c)^{3/2}}{6c}+\frac{(1-4c)^{3/2}}{6c}\frac{1}{n}
-\log\left[\frac{1-2c-\sqrt{1-4c}}{1-2c+\sqrt{1-4c}}\right]\frac{D}{n}
+O(n^{-2})\]
and
\beqas
F(x_0)=&~&\frac{1-6c+6c^2-(1-4c)^{3/2}}{12c}-
\frac{1-6c-6c^2-(1-4c)^{3/2}}{12c}\frac{1}{n}+\\
&-&\log\left[n\frac{1-2c-\sqrt{1-4c}}{2c}\right]\frac{D}{n}+O(n^{-2}).
\eeqas
The function $F(x)$ is plotted in figure
\ref{fig3} for $c=0.1$ as a function of $D$.
The ``entropic'' contribution, $S(x)$ is of order $1/n$. 
This is a consequence of the fact that $B_{3,0}=0$ in this model. 
Since $B_{2,1}(y,y)=-2(1-y)/c<0$, fluctuations 
in $y_k$ increase the probability of the state $x_0$. 
Indeed for large enough $D$, figure \ref{fig3} shows that
the state at $x_0$ has a higher probability.
Therefore fluctuations in this case {\em decrease} the
probability of the Pareto dominant NE.
Note also that, as $D\to 0$ and $n\to\infty$, 
$F(x_0)<F(x_-)$ for $c>2/9=0.2222\ldots$, which
implies that the probability to find the system in the 
Pareto dominant NE
$x_-$ tends to $0$. This example shows that the system does
not always choose the state of maximum utility. 
In addition, when $D>0$ one stable state can become unstable.
In our example, for higher values of $D$ or $c$ the state at 
$x_-$ which is a minimum of $U(x)$ is no more a minimum of
$F(x)$. 

\begin{figure}
\centerline{\psfig{file=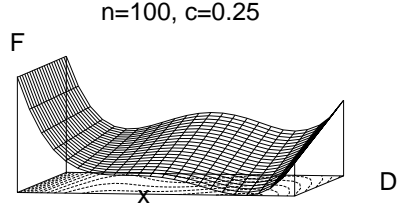,width=7cm,angle=270}}
%\vspace{5cm}
\caption{$F(x)$, as a function of $D$ for the 
utility function (\ref{b4}): $n=100$, $c=0.25$ and $b=0.1$.}
\label{fig4}
\end{figure}

In this system, however, entropic 
effects are ``accidentally'' small because $B_{3,0}=0$.
If one adds a higher order term the situation changes.
Consider indeed
\beq
B(x,y)=-x(1-y)\left[1-\frac{1-bc^3}{2c}x(1-y)-
\frac{b}{4}x^3(1-y)^3\right]\,,~~~~~c>0\,.
\label{b4}
\eeq
For $0<b<\frac{4}{c^3}$ this has the same stable equilibria as 
before\footnote{For $b<0$ the system is unstable and for
$b>4c^{-3}$, $x_-$ becomes unstable and two new equilibria appear.}. 
The picture, in the limit $n\to\infty$, is qualitatively
the same apart from the entropy, which now is finite 
since $B_{3,0}(x,x)=6bx(1-x)^4\neq 0$. Note that $B_{2,1}<0$,
and $B_{3,0}>0$. Therefore the effects of fluctuations are opposite to the 
ones discussed above. Indeed $S(x_0)<S(x_-)$, which means 
that the stochastic force favours, in this case, the Pareto dominant
NE $x_-$ over the state $x_0$. Contrary to our intuition from 
statistical mechanics, it is the ``ordered'' state $x_-$ which
has a larger contribution from the stochastic term.
This is shown in figure \ref{fig4}, which shows in particular that
fluctuations lead to a new minimum of $F(x)$: This 
meta-stable state is a precursor of the NE at $x_-$.
This example shows that the identification of $S(x)$ with an
entropy can be misleading. 

Direct numerical simulations of the Langevin equation
gives results in good agreement with this picture for
small values of $D$. The higher order terms in the $D$
expansion seem to enhance the behavior discussed above
for the two particular models.

For a particle in a random potential $F(\bar x)$
subject to a stochastic force of strength $T=D/n+D_0$, 
the transition times from one metastable state to the 
other are of the order of $\tau\approx\exp\{n[F(x_i)-F(x_+)]/
(D+nD_0)\}$ where $i=0,~-$ labels the state of departure. 
The generalization of this result to our 
case, predicts relaxation times that, for $D_0=0$
diverge in the ``thermodynamic'' limit $n\to \infty$.
This behavior is reminiscent of a first order phase
transition in statistical mechanics. 
It is worth to remind, however, that the transition 
from one state to the other is a far from equilibrium
process, whereas we derived eq. 
\req{fx} within an approximation which is valid to order 
$D$ close to the equilibria (see eq. \ref{assump}).
For this reason we performed numerical simulations 
of the above bi-stable systems. Even though a systematic 
quantitative computation of transition times was too
demanding, we definitely found that numerical
simulations are in qualitative agreement with the
picture emerging from the $O(D)$ approximation.

\section{Conclusions}

We have introduced a simple stochastic dynamics for game theory. 
This assumes ``local'' rationality since any player tries to climb
the gradient of his utility function. This deterministic process 
is affected by a stochastic force which represents deviation
from rationality in the form of a ``heat bath''. We focused on
particular games with a global interaction which is typical
of socio-economic systems: each player's utility
depends on his strategy as well as on a global quantity.
The stable states of this dynamics coincide with the NE.
We studied the gaussian fluctuations around these NE and established 
that competitive equilibria are characterized by large fluctuations
which grow with the number of players. Large fluctuations 
imply great inequalities in the distribution of utility among 
players. Uneven distribution of goods is, unfortunately, 
very common in the economic world. Our analysis suggests
that this is related to the competitive nature of the Nash 
equilibrium. Players competing for a common resource have
broadly distributed utilities whereas players whose 
strategy is bounded by a direct utility loss (as in the
tragedy of commons with taxes) have more or less the
same payoffs. Fluctuations usually
decrease the utility of players, but cases where the contrary holds
are also possible. Finally we studied the general case in a small 
noise limit. We found that, depending on the particular,
game, fluctuations can either increase or decrease the dominance 
of a Pareto dominant state and that new metastable states can occur.

This approach can naturally be extended to games with a discrete
strategy space. For these, the Langevin dynamics can be replaced 
by e.g. Metropolis dynamics where each player tries to minimize
his own cost function.

Fluctuations and deviation from rationality are inevitable in the
real world. Reducing their strength $D_i$ costs time and money. 
This suggests a generalization of our work where $D_i$ is considered
as a parameter in the strategy space. If one then assumes that
players have ``local'' rationality so that the best thing they can 
do is to climb their utility gradient, one is left with a 
system where the strategy of each player consists in the choice
of the strength of the fluctuations $D_i$ and the rate $\Gamma_i$ 
at which they climb the potential. In terms of these parameters
$(\vec D,\vec\Gamma)$ we can define a generalized utility
function 
\beq
U_i(\vec D,\vec\Gamma)=\avg{u_i(x_i,x_{-i})|\vec D,
\vec\Gamma}+U_0(D_i,\Gamma_i)\,,
\eeq
where the first term is the average utility discussed in the
body of the paper. The second term is instead related to the
cost of achieving a noise reduction to strength $D_i$ and 
a rate of convergence $\Gamma_i$. 
In general we expect $U_0$ to be a decreasing function of $D_i$.
Furthermore infinite precision $D_i=0$ most likely requires 
an infinite cost. A possible candidate for $-U_0$ is
the entropy $U_0(D)\propto\avg{\log P(\eta)}=\log\sqrt{D}+C$.
In general we found that the average utility 
decreases with increasing $D$. In these 
cases Nash equilibria, in the strategy space $(\vec D,\vec \Gamma)$ 
will occur for $D_i>0$. In the particular cases where we found that
$\avg{u_i(x_i,x_{-i})}$ increases with $D_i$, a large noise strength 
will be preferred to a more rational behavior. This new approach 
would provide a more realistic description of real systems of 
interacting players. 

This work was partly supported by the Swiss National Foundation 
under grant 20-46918.96


\begin{thebibliography}{99}

\bibitem{pw} Anderson, P.W.,  K. Arrow, D. Pines,  (1988),
{\em The Economy as an evolving complex system}, 
Redwood city, Addison-Wesley Co.

\bibitem{scaling} 
Mandelbrot, B.B., (1963) {\it J. of Business}, {\bf 36}, 394;
Mantegna, R.N. and H.E. Stanley, (1995), {\it Nature}, {\bf 376}, 46;
Levy, M., H. Levy and S. Solomon, (1995), J. Physique {\bf I 5}, 1087;
Galluccio, S., G. Caldarelli, M. Marsili and Y.-C. Zhang, (1997)
Physica A, {\bf 245}, 423.

\bibitem{multiscaling} Ghashghaie, S. {\it et al.}, (1996) {\it Nature}, 
{\bf 381}, 767.

\bibitem{soc} Bak, P. , M. Paczuski and M. Shubik, (1996) preprint 
cond-mat/9609144 (Submitted to the Journal of Mathematical Economics).

\bibitem{model} Caldarelli, G., M. Marsili and Y.-C. Zhang, (1997)
Europhys. Lett. {\bf 40}, 479.

\bibitem{games} Fudenberg D. and J. Tirole, (1991), {\em Game 
Theory}, The MIT Press.

\bibitem{nash} Nash, J., (1950), Proc. Nat. Acad. Sci. {\bf 36}, 48.

\bibitem{fluct} Marsili, M. and Y.-C. Zhang, (1997),
Physica A, {\bf 245}, 181.

\bibitem{pareto} Pareto, V. (1965),
{\em Ecrits sur la courbe de la r\'epartition
de la richesse}, ed. G. Busino (Droz, Gen\'eve). 

\bibitem{courn} Cournot, A. (1897) {\em Researches into the 
Mathematical Principles of the Theory of Wealth}, Ed. 
N. Bacon, New York Macmillan.

\bibitem{hardin} Hardin, G. (1968), {\em Science} {\bf 162}, 
1243.

\bibitem{evol}  Weibull, J.W., (1995), {\em Evolutionary Game 
Theory}, The MIT Press, (Cambridge, Massachusetts;
London, UK.

\bibitem{trembl} Selten, R., (1975) Int. J. of Game Th. {\bf 4} 25,
studied the robustness of a NE with respect to players' 
{\em trembling hands}. Trembling hand means that players are 
unable to play exactly pure strategies: any strategy is played
with a probability that cannot be less than a small number 
$\epsilon$. If a NE, in spite of this, does not change
dramatically, then it is said to be perfect.

\bibitem{wujiang} Wu, W., and J. Jiang, (1962), Sci. Sinica {\bf 11}, 
1307, investigated robustness of a NE with respect to small 
random perturbations in the payoff functions\cite{wujiang}.

\bibitem{maynard} Maynard Smith, J. and G. R. Price, 
Nature, {\bf 246}, 15 (1973); see also Maynard Smith, J. 
%J. Theor. Biol. {\bf 47}, 209 (1974), 
{\em Evolution and the theory of Games},
Cambridge Univ. Press (Cambridge, 1982).

\bibitem{repl} Taylor, P., and L. Jonker, (1978), Math. Biosc. 
{\bf 40}, 145; for a review, see J. Hofbauer and K. Sigmund, 
{\em The Theory of Evolution and Dynamical Systems}, Cambridge 
Univ. Press (Cambridge, 1988).

\bibitem{FYFH} Foster, D., and P. Young, (1990), Theor. Pop. Biol. 
{\bf 38}, 219; D. Fudenberg and C. Harris, J. Econ. Theory,
{\bf 57}, 420 (1992).

\bibitem{ref} Hohenberg, P.C., and B. I. Halperin, (1977), Rev. 
Mod. Phys. {\bf 49}, 435.
 
\bibitem{gardiner} 
Gardiner, C. W., (1985), {\em Handbook of Stochastic Methods}, 
second edition, p. 124 (Springer-Verlag). 

\end{thebibliography}
\end{document}